*Article*

# A Multiple Linear Regression Analysis to Measure the Journal Contribution to the Social Attention of Research

Pablo Dorta-González [1,*]

1  Institute of Tourism and Sustainable Economic Development (TIDES), University of Las Palmas de Gran Canaria, Campus de Tafira, 35017 Las Palmas de Gran Canaria, Spain
*  Correspondence: pablo.dorta@ulpgc.es

**Abstract:** This paper proposes a three-year average of social attention as a more reliable measure of social impact for journals, since the social attention of research can vary widely among scientific articles, even within the same journal. The proposed measure is used to evaluate a journal's contribution to social attention in comparison to other bibliometric indicators. The study uses Dimensions as a data source and examines research articles from 76 disciplinary library and information science journals through multiple linear regression analysis. The study identifies socially influential journals whose contribution to social attention is twice that of scholarly impact as measured by citations. In addition, the study finds that the number of authors and open access have a moderate impact on social attention, while the journal impact factor has a negative impact and funding has a small impact.

## 1. Introduction

Several factors influence the relationship between the social attention of a paper and its scholarly impact. Academic impact is primarily influenced by the perceived quality of the research, the reputation of the authors, their institutions, and the journals in which they are published. Social attention, on the other hand, is influenced by a broader range of factors, including the topic of the publication, the demographics of the authors who are active on social media, and the current trends and interests of the general public. For example, topics that are controversial or fashionable tend to generate a lot of social attention, regardless of the academic impact of the publication.

Ever since the term 'altmetrics' was first coined in 2010 [2], both theoretical and practical research has been carried out in this field [3]. Additionally, governments are now encouraging researchers to engage in activities that have social impact, such as those that bring economic, cultural, and health benefits [1].

While altmetric data can enhance citation rates by accelerating the accumulation of citations after publication [4], they only show a moderate correlation with Mendeley readership [5] and a weak or negligible correlation with other altmetric indicators and citations [6,7]. Consequently, altmetrics can capture various types of impact beyond citation impact [8].

There are many and varied factors that influence the public exposure of research [3]. These include collaboration, research funding, type of access to the publication, citation, and journal impact factor, which will be discussed below.

The importance of collaboration in scientific research is growing as it enables the combination of knowledge and skills to generate new ideas and research avenues [9]. Co-authorship analysis of research papers is a valid method for examining collaboration [10]. While researchers have increased their production of research articles in recent decades, the number of co-authors has also risen, resulting in a steady publication rate [11,12]. Collaborative research has been linked to higher citation rates and more impactful science [13]. Scientists who collaborate more frequently tend to have higher h-indices [14]. Additionally, both citation and social attention increase with co-authorship, though the influence becomes less significant as the number of collaborators increases [15].

Funding is another important input into the research process. Authors found that 43% of the publications acknowledge funding, with considerable variation across countries [16]. They also found that publications that acknowledge funding are more highly cited. However, citations are only one side of a multidimensional concept such as research impact, and alternatives have been considered to complement the impact of a research. Other authors conclude that there is a positive correlation between funding and usage metrics, but with differences between disciplines [17].

Another factor to consider when analyzing research performance is the type of access to the publication. The impact advantage of open access is probably because greater access allows more people to read articles that they would not otherwise read. However, the true causality is difficult to establish because there are many possible confounding factors [18,19].

In this context, in the present paper I quantify the contribution of the journal to the social attention to research. I compare this contribution with other bibliometric and impact indicators discussed earlier, such as collaboration, research funding, type of access, citation, and journal impact factor. To this end, I propose a measure of social influence for journals. This indicator is a three-year average of the social attention given to articles published in the journal. The data source is *Dimensions* and the units of study are research articles in Library and Information Science. The methodology used is a multiple linear regression analysis.

## 2. Social Attention to Research and Traditional Metrics

Scholars have been exploring the social impact of research papers in the field of research evaluation. However, traditional metrics like citation analysis, impact factors, and h-index tend to focus solely on the academic use of research papers and ignore their social impact on the internet [20]. Web 2.0 has transformed social interaction into a web-based platform that allows two-way communication and real-time interaction, creating an environment in which altmetrics have emerged as a new metric to measure the academic impact of a research paper. The term "altmetrics" was first introduced in 2010 by Priem, who also published a manifesto on the subject [2]. However, the correlation between citation counts and alternative metrics is complicated because neither are direct indicators of research quality, making it nearly impossible to achieve a perfect correlation unless they are unbiased [21].

Some research studies have focused on exploring the correlation between traditional citation metrics and alternative metrics [21,22]. Such studies are important for understanding how research performance is evaluated, especially in terms of measuring impact through citation counts and altmetric attention scores (AAS). While citation counts have been the primary means of assessing research performance, the importance of AAS is increasing in today's social media-driven world. This is because citation counts have limitations, such as delays in a publication being added to citation databases and potential biases arising from self-citations.

A research study compared citation data from three databases (WoS, Scopus, and Google Scholar) for 85 LIS journals, and found that citation data from Google Scholar had a strong correlation with altmetric attention, while the other two databases showed only a moderate correlation [23]. However, for the nine journals that were consistently present in all three databases, there was a positive but not significant correlation between citation score and altmetric attention. Although there was no correlation between citation count and altmetric score, the study found a moderate correlation between journal impact factor and citation count, a weak correlation between journal tweets and impact factor, and a strong correlation between journal tweets and altmetric score [24].

## 3. Materials and Methods

This study employed a rigorous and systematic process for collecting and analyzing bibliometric data using the Dimensions database to ensure that the results were reliable, valid, and informative for the field of library and information science.

- Unit of Analysis: The unit of analysis for this study is the "research article" in the field of library and information science.
- Data Source: The data source for this study is the Dimensions database, which provides social attention data at the article level. This database was selected because of its comprehensive coverage of scholarly publications in a variety of disciplines.
- Journal Selection: The JCR Journal Impact Factor in the Web of Science database was used to select journals in the library and information science category. Of the 86 journals identified in the 2020 edition, 10 journals were excluded because they were not indexed in the Dimensions database. This step ensured that only high-quality, peer-reviewed journals were included in the analysis.
- Final Dataset: The final dataset included 76 library and information science journals. These journals were selected based on their relevance and impact in the field.
- Timeframe: Research articles indexed in the Dimensions database between 2012 and 2021 were included in the analysis. This timeframe ensured that the analysis covered recent publications, while also allowing for enough data to be collected.
- Total number of articles: A total of 49,202 research articles were analyzed in this study. These articles were retrieved using the Dimensions database search function, which allowed for advanced filtering and sorting options.
- Data Collection: Data were collected on June 6, 2022, using the export function of the Dimensions database. The export file contained data on each research article's bibliographic information, social attention metrics, and other relevant variables.

Note that Altmetric is the source of altmetric data in Dimensions, and it is one of the earliest and most popular altmetric aggregator platforms. Digital Science launched this platform in 2011, and it tracks and accumulates mentions and views of scholarly articles from different social media channels, news outlets, blogs, and other platforms. It also computes a weighted score named the 'altmetric attention score', in which each mention category contributes differently to the final score [25].

The metric known as the altmetric attention score gauges the level of social attention an article receives from sources such as mainstream and social media, public policy documents, and Wikipedia. It assesses the article's online presence and evaluates the discussions surrounding the research. In this paper, to avoid confusion, the term "social attention score" or simply "social attention" is used to refer to this metric.

In this paper, a journal-level measure of social attention to research is proposed. This measure is defined as the average social attention of articles over a three-year time window. Note that the *Dimensions* database does not provide journal-level impact indicators. Therefore, I included another measure of journal impact in the dataset. I used the Journal Impact Factor provided by the JCR *Web of Science* database for 2020, the year available at the time of data collection.

The methodology consists of a multiple linear regression analysis. Thus, the dependent variable is the social attention of the article and the independent variables are the proposed measure of the social attention of the journal, the number of authors, the type of access to the publication {open access = 1, closed = 0}, the funding of the research {funded = 1, unfunded = 0}, the citations of the article and the impact factor of the journal.

## 4. Results

The article-level dataset is described in Table 1. The information in this table is presented according to the time elapsed since publication (in average years from publication to the time of data collection in the first half of 2022). It can be observed that the maximum social attention of a scientific research in library and information science is reached on average four years after its publication, with an average score of 5.57. However, there are no significant differences after the second year. The highest values, more than five points, are observed between the second and sixth year after publication. Nevertheless, the marginal variation between years is only relevant in the second year, with an increase of 0.97 points compared to the first year.

**Table 1.** Descriptive statistics of the dataset at the article level. Category Library and Information Science. Data Source Dimensions

| Years since Pub.* | Year of Pub. | Num. Art. | Num. Authors (Mean) | OA Art. (%) | Funded Art. (%) | Citations (Mean) | Art. Social Attention Mean Score | Art. Social Attention Marg. Var. |
|---|---|---|---|---|---|---|---|---|
| 1 | 2021 | 6,156 | 3.15 | 39.57% | 23.93% | 2.79 | 4.41 | |
| 2 | 2020 | 5,687 | 3.10 | 44.43% | 26.01% | 9.14 | 5.39 | 0.97 |
| 3 | 2019 | 4,880 | 2.88 | 43.03% | 23.55% | 11.87 | 5.31 | -0.08 |
| 4 | 2018 | 4,811 | 2.73 | 43.17% | 22.49% | 14.94 | 5.57 | 0.26 |
| 5 | 2017 | 4,897 | 2.63 | 45.01% | 21.14% | 16.91 | 5.17 | -0.40 |
| 6 | 2016 | 4,592 | 2.72 | 42.29% | 18.47% | 20.61 | 5.51 | 0.34 |
| 7 | 2015 | 4,521 | 2.64 | 39.02% | 20.97% | 23.14 | 4.92 | -0.58 |
| 8 | 2014 | 4,551 | 2.67 | 37.84% | 19.12% | 22.73 | 3.89 | -1.03 |
| 9 | 2013 | 4,613 | 2.45 | 34.92% | 17.86% | 24.36 | 3.33 | -0.57 |
| 10 | 2012 | 4,494 | 2.37 | 34.40% | 15.20% | 25.77 | 2.57 | -0.75 |
| All | | 49,202 | 2.76 | 40.50% | 21.12% | 16.52 | 4.63 | |

* Average years from publication to the time of data collection in the first half of 2022

Table 1 also shows how the average number of authors per article in library and information science has gradually increased over the past decade, from an average of 2.37 authors per article in 2012 to an average of 3.15 authors per article in 2021. This 33% increase in co-authorship in a decade is remarkable. The increase in co-authorship may partly explain the 37% increase in research article production over the decade in the Library and Information Science category, from just under 4,500 articles in 2012 to more than 6,100 articles in 2021.

In the dataset, 40% of the articles are open access, and 21% of the publications indicate in the acknowledgments section that the authors have received some form of funding, with a sustained increase in most of the years analyzed. In terms of citations, the increase observed in Table 1 was to be expected, from 2.8 cites per article in the first year after the publication to an average of 25.8 cites at the end of the decade. Significant marginal increases are observed up to the seventh year after publication, highlighting the increase of 6.35 citations that occurs in the second year.

### 4.1. Journal Social Attention: Definition and Consistency of the Indicator

When aggregating the data at the journal level, I observed a large interannual variability in the average social attention per article when the time window is reduced to a single year. That is, for each journal, the average social attention of the articles in a given year differs significantly from that of the articles in the previous and subsequent years of the series. This large variability in the average social attention of each journal over time means that the one-year average is not a consistent measure of social attention for journals. This weakness observed for social media mentions also occurs for other citation-based indicators, such as the impact factor, with short time windows.

One reason for this high inter-annual variability is the low correlation between the individual scores of articles and the average scores of journals when the time window in which citations or mentions are collected is short. Thus, a small proportion of articles from each journal receive a large proportion of the scientific citations and social mentions. In order to increase the consistency of a measure by partially reducing the interannual variability, it is often chosen in bibliometrics to increase the size of the window for counting observations (citations or mentions). In the case of the impact factor, the various databases thus provide indicators for 2, 3, 4, and even 5 years.

In this study, I chose a three-year window as a compromise between the advantages and disadvantages of considering large time windows. That is, 4-year and 5-year windows require a long waiting period before social attention can be measured for a journal, while a two-year window still produces a high interannual variability in the dataset.

Therefore, I propose the following definition for the measure of social attention at the journal level. The journal social attention in year $y$ counts the social attention received in years $y$-2, $y$-1, and $y$ for research articles published in those years ($y$-2, $y$-1, and $y$) and divides it by the number of research articles published in those years ($y$-2, $y$-1, and $y$). For example, the journal social attention in 2021 counts the social attention received in 2019-2021 for research articles published in 2019-2021 and divides it by the number of research articles published in 2019-2021.

The journal-level dataset is described in Table 2. This table also includes the measure of the journal social attention. Due to space limitations, I only show the information corresponding to the year 2020 for the production and impact indicators (the last year available at the time of data collection), and the year 2021 for the journal social attention (time window 2019-2021). The graphical description of the data is shown in Figures 1 and 2.

**Table 2.** Description of the dataset at the journal level in the subject category Library and Information Science (JCR). Data Sources Web of Science and Dimensions

| | Journal | Num. Art. 2020 | JIF 2020 | JIF Percentile 2020 | JIF Quartile 2020 | 5-Year JIF 2020 | Journal Social Attention 2021 |
|---|---|---|---|---|---|---|---|
| 1 | ASLIB J INFORM MANAG | 54 | 1.903 | 44.12 | Q3 | 2.343 | 1.83 |
| 2 | CAN J INFORM LIB SCI | 5 | 0.000 | 0.59 | Q4 | 0.420 | 0.78 |
| 3 | COLL RES LIBR | 52 | 2.381 | 52.35 | Q2 | 2.204 | 5.55 |
| 4 | DATA TECHNOL APPL | 51 | 1.667 | 39.41 | Q3 | 1.667 | 0.37 |
| 5 | ELECTRON LIBR | 56 | 1.453 | 34.71 | Q3 | 1.540 | 0.36 |
| 6 | ETHICS INF TECHNOL | 64 | 4.449 | 74.71 | Q2 | 3.925 | 10.70 |
| 7 | EUR J INFORM SYST | 64 | 4.344 | 71.18 | Q2 | 7.130 | 6.43 |
| 8 | GOV INFORM Q | 71 | 6.695 | 91.18 | Q1 | 8.293 | 7.66 |
| 9 | HEALTH INFO LIBR J | 45 | 2.154 | 47.65 | Q3 | 2.187 | 6.25 |
| 10 | INFORM SOC-ESTUD | 32 | 0.311 | 8.82 | Q4 | 0.313 | 0.39 |
| 11 | INFORM MANAGE-AMSTER | 91 | 7.555 | 94.71 | Q1 | 9.183 | 2.48 |
| 12 | INFORM ORGAN-UK | 15 | 6.300 | 90 | Q1 | 5.866 | 6.24 |
| 13 | INFORM DEV | 74 | 2.049 | 46.47 | Q3 | 2.205 | 0.99 |
| 14 | INFORM PROCESS MANAG | 237 | 6.222 | 88.82 | Q1 | 5.789 | 2.06 |
| 15 | INFORM RES | 75 | 0.780 | 20.59 | Q4 | 1.197 | 1.42 |

| | Journal | Num. Art. 2020 | JIF 2020 | JIF Percentile 2020 | JIF Quartile 2020 | 5-Year JIF 2020 | Journal Social Attention 2021 |
|---|---|---|---|---|---|---|---|
| 16 | INFORM SOC | 25 | 4.571 | 77.06 | Q1 | 3.936 | 6.94 |
| 17 | INFORM SYST J | 49 | 7.453 | 93.53 | Q1 | 8.814 | 7.05 |
| 18 | INFORM SYST RES | 69 | 5.207 | 82.94 | Q1 | 6.888 | 7.54 |
| 19 | INFORM TECHNOL MANAG | 15 | 1.533 | 38.24 | Q3 | 2.627 | 0.19 |
| 20 | INFORM TECHNOL PEOPL | 121 | 3.879 | 67.65 | Q2 | 4.238 | 1.65 |
| 21 | INFORM TECHNOL LIBR | 27 | 1.160 | 27.65 | Q3 | 1.351 | 8.90 |
| 22 | INFORM TECHNOL DEV | 66 | 4.250 | 70 | Q2 | 4.221 | 6.06 |
| 23 | INT J COMP-SUPP COLL | 18 | 5.108 | 80.59 | Q1 | 4.966 | 4.09 |
| 24 | INT J GEOGR INF SCI | 162 | 4.186 | 68.82 | Q2 | 4.645 | 3.66 |
| 25 | INT J INFORM MANAGE | 203 | 14.098 | 99.41 | Q1 | 13.074 | 4.99 |
| 26 | INVESTIG BIBLIOTECOL | 39 | 0.475 | 13.53 | Q4 | 0.535 | 0.27 |
| 27 | J ACAD LIBR | 104 | 1.533 | 38.24 | Q3 | 2.023 | 5.03 |
| 28 | J COMPUT-MEDIAT COMM | 25 | 5.410 | 85.29 | Q1 | 9.953 | 25.01 |
| 29 | J DOC | 98 | 1.819 | 40.59 | Q3 | 1.988 | 3.43 |
| 30 | J ENTERP INF MANAG | 128 | 5.396 | 84.12 | Q1 | 5.839 | 0.26 |
| 31 | J GLOB INF MANAG | 38 | 1.373 | 33.53 | Q3 | 1.550 | 0.75 |
| 32 | J GLOB INF TECH MAN | 14 | 3.519 | 66.47 | Q2 | 2.631 | 0.94 |
| 33 | J HEALTH COMMUN | 94 | 2.781 | 59.41 | Q2 | 3.468 | 8.44 |
| 34 | J INF SCI | 120 | 3.282 | 65.29 | Q2 | 2.904 | 2.38 |
| 35 | J INF TECHNOL-UK | 21 | 5.824 | 86.47 | Q1 | 9.439 | 2.28 |
| 36 | J INFORMETR | 77 | 5.107 | 79.41 | Q1 | 5.421 | 7.12 |
| 37 | J KNOWL MANAG | 162 | 8.182 | 97.06 | Q1 | 8.720 | 0.60 |
| 38 | J LIBR INF SCI | 123 | 1.992 | 45.29 | Q3 | 2.009 | 2.47 |
| 39 | J MANAGE INFORM SYST | 40 | 7.838 | 95.88 | Q1 | 8.335 | 3.53 |
| 40 | J ORGAN END USER COM | 23 | 4.349 | 72.35 | Q2 | 2.808 | 0.08 |
| 41 | J SCHOLARLY PUBL | 19 | 1.512 | 35.88 | Q3 | 1.245 | 3.77 |
| 42 | J STRATEGIC INF SYST | 16 | 11.022 | 98.24 | Q1 | 11.832 | 1.37 |
| 43 | J AM MED INFORM ASSN | 209 | 4.497 | 75.88 | Q1 | 5.178 | 14.38 |
| 44 | J ASSOC INF SCI TECH | 150 | 2.687 | 54.71 | Q2 | 3.854 | 8.09 |
| 45 | J ASSOC INF SYST | 47 | 5.149 | 81.76 | Q1 | 6.780 | 2.06 |
| 46 | J AUST LIB INF ASSOC | 28 | 0.725 | 19.41 | Q4 | 0.851 | 1.64 |
| 47 | J MED LIBR ASSOC | 60 | 3.180 | 61.76 | Q2 | 3.874 | 4.57 |
| 48 | KNOWL MAN RES PRACT | 106 | 2.744 | 58.24 | Q2 | 3.027 | 1.24 |
| 49 | KNOWL ORGAN | 34 | 1.000 | 25.29 | Q4 | 0.979 | 0.18 |
| 50 | LEARN PUBL | 53 | 2.506 | 53.53 | Q2 | 2.659 | 20.19 |
| 51 | LIBR INFORM SCI RES | 32 | 2.730 | 57.06 | Q2 | 2.778 | 5.15 |
| 52 | LIBR HI TECH | 83 | 2.357 | 51.18 | Q2 | 2.065 | 1.32 |
| 53 | LIBR QUART | 26 | 1.895 | 42.94 | Q3 | 2.277 | 1.27 |
| 54 | LIBR RESOUR TECH SER | 3 | 0.424 | 12.35 | Q4 | 0.541 | 0.45 |
| 55 | LIBR TRENDS | 38 | 1.311 | 31.18 | Q3 | 1.354 | 5.70 |

| | Journal | Num. Art. 2020 | JIF 2020 | JIF Percentile 2020 | JIF Quartile 2020 | 5-Year JIF 2020 | Journal Social Attention 2021 |
|---|---|---|---|---|---|---|---|
| 56 | LIBRI | 25 | 0.521 | 14.71 | Q4 | 0.706 | 0.75 |
| 57 | MALAYS J LIBR INF SC | 20 | 1.250 | 28.82 | Q3 | 1.320 | 0.00 |
| 58 | MIS QUART | 58 | 7.198 | 92.35 | Q1 | 12.803 | 0.84 |
| 59 | MIS Q EXEC | 17 | 4.371 | 73.53 | Q2 | 7.563 | 6.12 |
| 60 | ONLINE INFORM REV | 82 | 2.325 | 50 | Q3 | 2.883 | 3.92 |
| 61 | PORTAL-LIBR ACAD | 36 | 1.067 | 26.47 | Q3 | 1.285 | 1.93 |
| 62 | PROF INFORM | 169 | 2.253 | 48.82 | Q3 | 2.285 | 6.99 |
| 63 | QUAL HEALTH RES | 192 | 3.277 | 64.12 | Q2 | 5.038 | 7.03 |
| 64 | REF USER SERV Q | 27 | 0.650 | 17.06 | Q4 | 0.581 | 0.22 |
| 65 | REF SERV REV | 38 | 0.831 | 22.94 | Q4 | 1.221 | 1.74 |
| 66 | RES EVALUAT | 20 | 2.706 | 55.88 | Q2 | 3.434 | 12.70 |
| 67 | RESTAURATOR | 13 | 0.296 | 7.65 | Q4 | 0.427 | 0.11 |
| 68 | REV ESP DOC CIENT | 29 | 1.276 | 30 | Q3 | 1.259 | 2.71 |
| 69 | SCIENTOMETRICS | 454 | 3.238 | 62.94 | Q2 | 3.702 | 7.97 |
| 70 | SERIALS REV | 36 | 0.324 | 10 | Q4 | 0.425 | 1.53 |
| 71 | SOC SCI COMPUT REV | 93 | 4.578 | 78.24 | Q1 | 5.194 | 11.99 |
| 72 | SOC SCI INFORM | 28 | 0.714 | 18.24 | Q4 | 0.966 | 4.48 |
| 73 | TELECOMMUN POLICY | 84 | 3.036 | 60.59 | Q2 | 3.500 | 9.72 |
| 74 | TELEMAT INFORM | 91 | 6.182 | 87.65 | Q1 | 6.769 | 3.61 |
| 75 | TRANSINFORMACAO | 23 | 0.648 | 15.88 | Q4 | 0.561 | 5.20 |
| 76 | Z BIBL BIBL | 19 | 0.125 | 1.76 | Q4 | 0.071 | 0.31 |

A box-and-whisker plot by quartile for the 5-year journal impact factor (2020 edition) and the journal social attention (2021 edition) is shown in Figure 1. The differences between the groups are all significant for the journal social attention (p < 0.01), except between Q1 and Q2. Note that the journal social attention decreases in groups Q2 to Q4 as journals reduce their impact factors. This trend is observed in both the mean and the median, and even in the remaining quartiles of the distribution represented by the boxes and whiskers in the figure. However, this is not the case when moving from Q1 to Q2.

As can also be seen in Figure 1, the distributions in the group of journals with a higher impact factor (Q1) are highly skewed, especially in terms of social attention. Note that the mean, represented by the cross, is much higher than the median, represented by the central line in the box.

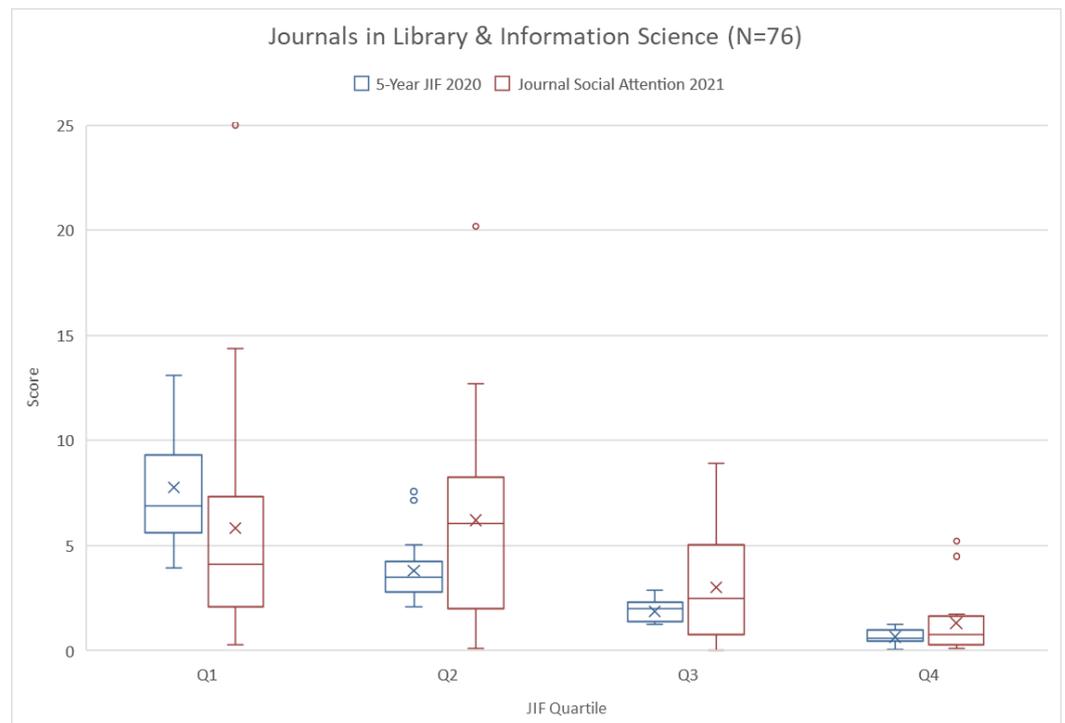

**Figure 1.** The 5-year Journal Impact Factor (edition 2020) and the Journal Social Attention (edition 2021) by JIF quartiles. The mean is represented by a cross. The differences between the groups are all significant ($p < 0.01$), except between Q1 and Q2 for the journal social attention.

Figure 2 shows the enormous variability in social attention regardless of the journal's impact factor. Note that in the group of medium-impact journals (Q2 and Q3) there are articles that receive a lot of social attention.

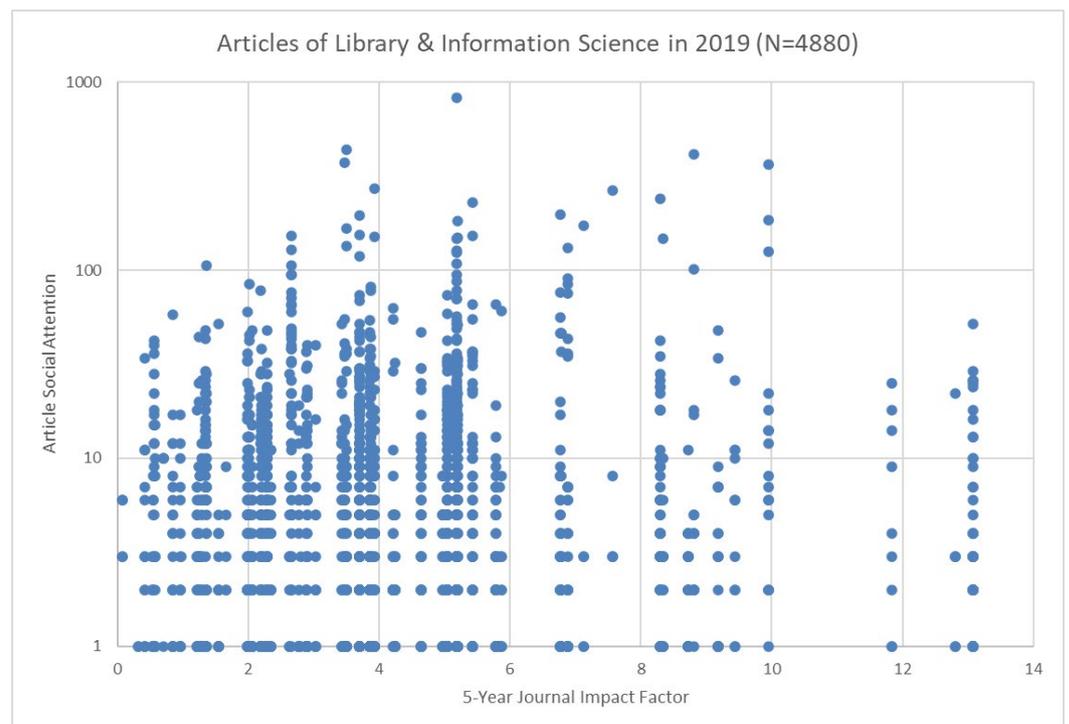

**Figure 2.** Social attention (logarithmic scale) of articles published in 2019, in relation to the journal's 5-year impact factor (2020 edition), in the category Library & Information Science.

*4.2. Multiple Linear Regression Analysis*

I would like to know if and how the social attention of articles can be predicted from the social attention of journals and several bibliometric characteristics. The description of the variables can be found in Table 3.

**Table 3.** Variables and description in the Multiple Linear Regression model.

| Name | Measure | Variable Description |
| --- | --- | --- |
| Article Social Attention | Altmetric attention score for research articles published in 2019 | Natural number $N = \{0, 1, 2, ...\}$ |
| Journal Social Attention | Journal social attention 2021: average altmetric attention score for research articles published in 2019-2021 | Positive real number $R^+_0 = [0, +\infty)$ |
| Num. Authors | Number of authors for research articles published in 2019 | Positive natural number $N^+ = \{1, 2, 3, ...\}$ |
| OA Article | Type of access to the research articles published in 2019 | Dichotomous {open access = 1, closed = 0} |
| Funded Article | Research funding declared in the acknowledgments section for the research articles published in 2019 | Dichotomous {funded = 1, unfunded = 0} |
| Article Citations | Times cited in Dimensions for research articles published in 2019 | Natural number $N = \{0, 1, 2, ...\}$ |
| Journal Impact Factor | 5-year journal impact factor in the 2020 issue of JCR | Positive real number $R^+_0 = [0, +\infty)$ |

The dependent variable is the article social attention for publications in the year 2019, in short "article social attention". The independent variables are the journal social attention (2021), the number of authors for publications in the year 2019, the type of access to these publications {open access = 1, closed = 0}, the funding of the research {funded = 1, unfunded = 0}, the article citations, and the journal impact factor (2020 edition).

I checked for missing values in the descriptive statistics of all variables (see Table 4 for the mean and standard deviation). Note that I have N = 4880 independent observations in the dataset. The distributions in the histograms are likely for all variables and there are no missing values. I have also checked for curvilinear relationships or anything unusual in the plot of the dependent variable against each independent variable.

Note that each independent variable has a significant linear relationship with the article social attention (see Table 4). Therefore, the multiple linear regression model could estimate the article social attention from all independent variables simultaneously. I checked the correlations between the variables (Table 4). Absolute correlations are low (none of the correlations exceeds 0.39) and multicollinearity problems are discarded for the actual regression analysis.

In general, the observed correlations are low. The highest correlations are between article citations and journal impact factor (0.39) and between number of authors and funding (0.33). The remaining correlations are below 0.22. The only negative correlation is observed between the type of access and the impact factor.

**Table 4.** Means, SDs and Pearson correlations between the dependent and all independent variables.

| Variable | Mean | SD | 1 | 2 | 3 | 4 | 5 | 6 | 7 |
|---|---|---|---|---|---|---|---|---|---|
| 1. Article Social Attention | 5.31 | 22.42 | - | .22** | .09** | .10** | .04** | .10** | .03* |
| 2. Journal Social Attention | 5.08 | 4.29 | | - | .20** | .19** | .20** | .03* | .15** |
| 3. Num. Authors | 2.88 | 2.10 | | | - | .07** | .33** | .11** | .19** |
| 4. OA Article (1, Closed = 0) | 0.43 | 0.50 | | | | - | .04** | .00 | -.11** |
| 5. Funded Article (1, Unfunded = 0) | 0.24 | 0.42 | | | | | - | .11** | .17** |
| 6. Article Citations | 11.87 | 26.62 | | | | | | - | .39** |
| 7. Journal Impact Factor | 4.24 | 3.03 | | | | | | | - |

*p < 0.05. **p < 0.01.

The regression model according to the b-coefficients in Table 5 is as follows:

$$\text{Article\_Social\_Attention}\_i = 0.74 \cdot \text{Journal\_Social\_Attention}\_i + 0.41 \cdot \text{Num\_Authors}\_i \\ + 1.75 \cdot \text{OA\_Article}\_i + 1.55 \cdot \text{Funded\_Article}\_i \\ + 0.09 \cdot \text{Article\_Citations}\_i - 0.78 \cdot \text{JIF}\_i \quad (1)$$

where Article_Social_Attention_$i$ denotes the predicted social attention for article $i$, $i=1,2,\ldots 4880$.

The R-squared is the proportion of the variance in the dependent variable accounted for by the model. I have reported the adjusted R-squared in Table 5. In this model $R^2_{adj}$ = 0.197. This is considered acceptable by social science standards. Furthermore, since the p-value found in the ANOVA is $p = 10^{-3}$, I conclude that the entire regression model has a non-zero correlation.

Note that each b-coefficient in equation (1) indicates the average increase in social attention associated with a one-unit increase in a predictor, all else equal. Thus, a 1-point increase in journal social attention is associated with a 0.74 increase in article social attention.

**Table 5.** Regression coefficients for the prediction of Article Social Attention. Standard Multiple Linear Regression analysis.

| Variable | B (Coeff.) | 95% CI | β (Standardized Coeff.) | t | p (Sig.) |
|---|---|---|---|---|---|
| Constant | 0 | - | 0 | - | - |
| Journal Social Attention | 0.741 | [0.595, 0.887] | 0.207 | 9.956 | 0.000 |
| Num. Authors | 0.412 | [0.215, 0.609] | 0.051 | 4.093 | 0.000 |
| OA Article (1, Closed = 0) | 1.755 | [1.040, 2.471] | 0.048 | 4.811 | 0.000 |
| Funded Article (1, Not = 0) | 1.552 | [0.358, 2.746] | 0.022 | 2.548 | 0.011 |
| Article Citation | 0.090 | [0.079, 0.101] | 0.104 | 16.373 | 0.000 |
| Journal Impact Factor | -0.784 | [-0.969, -0.600] | -0.044 | -8.336 | 0.000 |

Note. Adjusted R-square $R^2_{adj}$ = 0.197 (N=4880, p=$10^{-4}$). CI = confidence interval for B.

Similarly, an additional co-author contributes an average 0.41 increase in the social attention of the research. Furthermore, one additional citation increases the social attention of an article by an average of 0.09 points, or alternatively, every ten citations increases the social attention of a study by 0.9 points, all else being equal. Similarly, a one-point increase in the journal's impact factor is associated with a 0.78 decrease in the social attention of the article.

For the dichotomous variables, a 1 unit increase in open access is associated with an average 1.75 point increase in the social attention to the article. Note that open access is coded in the dataset as 0 (closed access) and 1 (open access). Therefore, the only possible 1 unit increase for this variable is from closed (0) to open (1). Therefore, I can conclude that the average social attention for open articles is 1.75 points higher than for closed articles (all other things being equal). Similarly, the average social attention for funded articles is 1.55 points higher than for unfunded articles, all else being equal.

The statistical significance column (Sig. in Table 5) shows the 2-tailed p-value for each b-coefficient. Note that all b-coefficients in the model are statistically significant ($p < 0.05$) and most of them are highly statistically significant with a p-value of $10^{-3}$. However, the b-coefficients do not indicate the relative strength of the predictors. This is because the independent variables have different scales. The standardized regression coefficients or beta coefficients, denoted as $\beta$ in Table 5, are obtained by standardizing all the regression variables before calculating the coefficients and are therefore comparable within and across regression models.

Thus, the two strongest predictors in the coefficients are the social attention of the journal ($\beta = 0.207$) and the citations received by the article ($\beta = 0.104$). This means that the journal is the factor that contributes the most to the social attention of the research, about twice as much as the citations received. In addition, the number of authors in the research ($\beta = 0.051$) contributes about half as much as citations and slightly more than open access to the publication ($\beta = 0.048$). Journal impact factor ($\beta = -0.044$) contributes as much as open access, but in the opposite direction. Finally, research funding ($\beta = 0.022$) contributes half as much as the impact factor.

Regarding the multiple regression assumptions, each observation corresponds to a different article. Therefore, I can consider them as independent observations. The regression residuals are approximately normally distributed in the histogram. I also checked the assumptions of homoscedasticity and linearity by plotting the residuals against the predicted values. This scatterplot shows no systematic pattern and therefore I can conclude that both assumptions hold.

## 5. Discussion

Social attention to research is crucial for understanding the impact and dissemination of scientific research beyond traditional citation-based metrics, and has practical implications for academic publishing, funding decisions, and science communication.

First, it provides a measure of the impact of scientific articles beyond traditional citation-based metrics, such as the number of times an article is shared, downloaded, or discussed on social media platforms. This can help researchers, publishers, and funding agencies better understand the impact and reach of their research. Second, social attention research can provide insights into how scientific information is disseminated and consumed by different audiences, which can inform public engagement and science communication strategies. Third, social attention research can highlight emerging trends and issues in science and technology that can guide future research agendas and funding decisions.

The results suggest that public attention to research occurs mainly in the first year after publication and to a lesser extent in the second year. However, a more detailed analysis of the dataset shows that the largest increase in social attention is observed in the first months after its publication.

Some considerations can be made about the negative signs observed in the interannual marginal variation (decrease in the average social attention compared to the previous year). This reduction may be due to several factors. First, the increasing use of social networks and the growing number of platforms from which social attention is collected. Second, because the social attention of research is measured with regularly updated data on social presence on the Internet (from June 2022 in the dataset). Since some mentions in social media may be ephemeral and disappear after a while (unlike citations in the databases that index the documents), the negative signs in the marginal variation could also be due to this circumstance. Finally, the observations (research articles) differ between years, so this result is therefore plausible.

The average number of authors per article in library and information science has gradually increased over the last decade, from an average of 2.37 authors per article in 2012 to an average of 3.15 authors in 2021. This 33% increase in co-authorship in a decade is relevant in terms of social attention, as discussion on the Internet is often driven by the authors of the research. More authors therefore mean more presence on social networks.

In the dataset, 40% of library and information science publications are open access. In addition, 21% of the publications indicate in the acknowledgments section that the authors have received some form of funding, with a sustained increase over the years. Note that this percentage of funded articles in LIS is half the average for all research fields [12].

The social attention of journals decreases in quartiles Q2 to Q4 as journals reduce their impact factors. However, there are no significant differences between the two highest quartiles (Q1 and Q2). This means that the journals that are most cited by researchers are not necessarily the ones that receive the most social attention. Note that this may be due to the subject category analyzed. For example, in the Library and Information Science category, there are also prestigious journals in the second quartile. This is the case, for example, for the journal *Scientometrics*. Journals with low obsolescence are penalized by the impact factor compared to other journals with higher obsolescence, which accumulate most of their citations in the first years after publication [26].

I found low correlations between the variables. The highest correlations are between citation count and journal impact factor (0.39) and between number of authors and funding (0.33). All other correlations are below 0.22. The only negative correlation is observed between the type of access and journal impact factor. This is because in the library and information science category, open access publishing is not yet widespread among the journals with the highest impact factors. Surprisingly, however, there is no correlation between access type and citations. In other words, open access articles do not receive more citations than closed articles. The reason for this is the same as that mentioned above. Open access in library and information science is not generalized in high-impact journals, which are those with the greatest visibility of research [27].

I observed that a one-point increase in the social attention of the journal is associated with an average 0.74 increase in the social attention of the article, all else equal. Similarly, an additional co-author contributes an average 0.41 increase in the social attention of the research. Furthermore, every ten citations increase the social attention of a paper by 0.9 points, all else being equal. I also concluded that the average social attention for open articles is 1.75 points higher than for closed articles, all else being equal. Similarly, the average social attention for funded articles is 1.55 points higher than for unfunded research.

The finding that a one-point increase in the journal impact factor is associated with a 0.78 decrease in the social attention to the article suggests that there is a negative relationship between the two metrics. One possible explanation for this result is that the number of citations, which is the basis of the journal impact factor, is an indicator of the influence of research in the academic world. The academic influence of a publication is determined by various factors, such as the reputation of the authors, the standing of the institutions they are affiliated with, and the perceived significance and quality of the research. Therefore, journals with high impact factors tend to publish research that is more specialized and may be of interest primarily to researchers in a particular topic, resulting in fewer social mentions.

However, social attention is affected by a wider range of factors, including the subject of the publication and the current trends and interests of the general public. For example, controversial or fashionable topics tend to generate a lot of social attention, regardless of the journal impact factor of the publication. Therefore, papers in high impact factor journals that do not address current social trends or controversial topics may not receive as much social attention as papers in low impact factor journals that address such topics.

Another factor that may contribute to the negative correlation between the journal impact factor and the social attention of a paper is the demographic of the authors who are active in social media. Younger researchers tend to be more active on social media than their more established counterparts, and they may have a higher likelihood of publishing in low-impact journals due to their less extensive research experience. This could also contribute to the negative correlation between the two metrics.

In summary, the negative association between the journal impact factor and the social attention of a paper can be explained by the different factors that influence the two metrics. While the journal impact factor is primarily influenced by academic factors such as the reputation of the authors and their institutions, the social attention of a paper is influenced by a wider range of factors, including the subject of the publication and the demographic of the authors.

The standardized regression coefficients indicate that the social attention of the journal and the citations received by the article are the two strongest predictors of the article's social attention. The analysis shows that the journal is the most influential factor, contributing approximately twice as much as the citations received. The number of authors in the research contributes roughly half as much as the citations, and slightly more than the publication being open access. The impact factor of the journal has a similar influence as open access, but in the opposite direction. Finally, research funding contributes about half as much as the impact factor.

There are some points to be noted regarding the use of hybrid indicators, and the "altmetric attention score" was used in this research. This is an example of a hybrid indicator that combines various sources to create a single score [25]. However, hybrid indicators are not robust, and hence should not be used to evaluate researchers, especially in hiring or internal promotions. In this study, the indicator was used to evaluate the research process, rather than the researchers themselves.

## 6. Conclusions

Understanding societal attention to research is important because it can help researchers identify emerging or pressing societal issues, prioritize research questions, and engage with stakeholders and the public. It can also inform efforts to communicate research findings to a broader audience, promote evidence-based policy, and increase public trust in science.

Although most of the social attention to research occurs in the first year, even in the first few months, a robust measure with low variability over time is preferable for identifying socially influential journals. This paper proposes a three-year average of social attention as a measure of social influence for journals. I used a multiple linear regression analysis to quantify the contribution of journals to the social attention of research in comparison to other bibliometric indicators. Thus, the data source was Dimensions, and the unit of study was the research article in disciplinary journals of Library and Information Science.

As a main result, the factors that best explain the social attention of the research are the social attention of the journal and the number of citations. However, there are socially influential journals, and their contribution to the social attention of the article multiplies by two the effect attributed to the academic impact measured by the number of citations. Furthermore, the number of authors and the open access have a moderate effect on the social attention of research. Funding has a small effect, while the impact factor of the journal has a negative effect.

It should be emphasized that low R-squared values may suggest that the forecasts of an article's social attention are not very accurate. Additionally, altmetric indicators offer the advantage of measuring various types of impact beyond scholarly citations, and they have the potential to identify earlier evidence of impact, making them valuable for self-assessment. Furthermore, they are useful in researching scholarship, as in this study. Nonetheless, it is critical to use social attention with care because it may present a limited and biased perspective of all forms of social impact.

In this study, a specific research area was analyzed over a given timeframe. However, to apply the findings to other areas, it is recommended to conduct further studies with more diverse data. As for future research directions, integrating characteristics of the authors - such as their research experience, h-index, affiliations, or social media presence - into the model could offer insights into the social impact of their research and its correlation with citations received. Furthermore, inclusion of these and other variables may enhance the R-squared of the model.

**Author Contributions:** Conceptualization, P.D.-G.; methodology, P.D.-G.; software, P.D.-G.; validation, P.D.-G.; formal analysis, P.D.-G.; investigation, P.D-G; resources, P.D.-G.; data curation, P.D.-G.; writing—original draft preparation, P.D.-G.; writing—review and editing, P.D.-G.; supervision, P.D.-G.

**Funding:** This research received no external funding.

**Data Availability Statement:** Data sourced from Dimensions, an inter-linked research information system provided by Digital Science (accessed on 6 June 2022).

**Acknowledgments:** Access to advanced features of the database allowed by Digital Science.

**Conflicts of Interest:** The authors declare no conflict of interest.